\documentclass{article}

\usepackage{arxiv}

\usepackage[utf8]{inputenc} 
\usepackage[T1]{fontenc}    
\usepackage{hyperref}       
\usepackage{url}            
\usepackage{booktabs}       
\usepackage{amsfonts}       
\usepackage{nicefrac}       
\usepackage{microtype}      
\usepackage{lipsum}		
\usepackage{graphicx}
\usepackage[square,numbers,sort&compress]{natbib}
\usepackage{doi}
\usepackage[capitalise]{cleveref}

\title{An investigation of fatigue damage growth in composites materials using the vibration response phase decay}

\author{ {Matias Lasen} \\
	Department of Engineering Technology\\
	University of Twente\\
	Enschede, 7522 NB, The Netherlands \\
	\texttt{m.a.lasenandrade@utwente.nl} \\
	\And
	{Dario Di Maio} \\
	Department of Engineering Technology\\
	University of Twente\\
	Enschede, 7522 NB, The Netherlands \\
	\texttt{d.dimaio@utwente.nl} \\
 	\And
	{Damaso De Bono} \\
	AtkinsRealis\\
 	\And
	{Michelle Peluzzo} \\
	Avio Aero\\
}

\renewcommand{\headeright}{}
\renewcommand{\undertitle}{}

\hypersetup{
pdftitle={An investigation of fatigue damage growth in composites materials using the vibration response phase decay},
pdfsubject={q-bio.NC, q-bio.QM},
pdfauthor={David S.~Hippocampus, Elias D.~Striatum},
pdfkeywords={First keyword, Second keyword, More},
}

\begin{document}
\maketitle

\begin{abstract}
	The increasing use of polymer composites in industry asks for the creation of better, faster and cost-effective methods to detect the damage state of such materials. This work presents the investigation of the phase decay , $\Delta\Phi$, as a new parameter to characterise crack growth in composites materials utilising an experimental framework of High Frequency Fatigue Testing (HFFT), a framework where the excitation occurs at vibration resonance. The proposed methodology empirically relates the crack growth measurements, from interrupted testing, with the structural phase decay response, distinctive of material strength degradation.
\end{abstract}


\section{Introduction}\label{Introduction}

Composite Fibre Reinforced Polymer (CFRP) materials are predominately used in structures where weight is critical, including aerostructures subjected to significant vibrations and cyclic loading. Due to the nature of cyclic loads, fatigue failure is a possibility and is therefore an important failure mode to understand. 

For CFRP materials there remains some uncertainty about phenomenological formulations for the crack driving force and crack resistance. The crack propagation rate is key to understanding fatigue behaviour. When this essential parameter is found, it can be used along Energy Release Rates (ERR), which is the change in potential (strain) energy resulting from an increase in crack area, to create empirical formulations relating applied loading to the crack propagation rate in a Paris law type of equations \cite{Davila2020}.  

The forces at the crack tip can lead to various crack opening modes, such as normal forces leading to opening mode (Mode-I), tangential forces to opening in-plane shearing mode (Mode-II) and a combination of both, leading to out-of-plane shearing mode (Mode-III), and the effect of mixing these modes influences the crack propagation rates and consequently the Paris Laws \cite{Allegri2013, Pascoe2013}. 

The main challenge in characterising the onset and growth of a crack is to measure the crack length accurately, which becomes more difficult for crack lengths below 2mm, because of the limited resolution of the measurement equipment \cite{Murray2018, Obaid2006, Roh2019, Panin2017}.

This paper embarks on the challenge of measuring very short cracks by proposing a completely different test than the one yielded by using Double Cantilever Beam technique. The testing techniques proposed in this manuscript are based on vibration tests, a sample is held by a fixture and vibrated under steady-state conditions at its first bending mode. The excitation frequency is consequently higher than usual standardised tests \cite{Maillet2013, Maillet2015}. The test methodology follows the experimental technique developed by Magi et al. and Voudouris et al. \cite{Magi2016,Voudouris2020}. The fundamental physical principle supporting the technique is measure the delta of the transfer function (ratio between stimulus and response) phase at a constant drive frequency and constant vibration amplitude. The transfer function phase $\phi(\omega)$ is very steep near the resonance, therefore, any variation in stiffness caused by the onset of damage is detected by measuring a change of phase, $\Delta\Phi$, over the number of excitation cycles. The phase curve over the number of cycles is a curve with negative slope, the decay of the phase tracks the change of the global stiffness of the sample subjected to fatigue test. 

The objective of this paper is to relate the variation of phase over cycles, $\Delta\Phi(N)$, to the crack length, $a$. To achieve that goal a two-step process is carried out, as explained in the Experimental Measurement (Section \ref{EM}). First, an empirical relation is found between phase decay and number of cycles, `$\Delta\Phi(N)$', in Section \ref{Res1}. The second step, in Section \ref{Res2}, builds the empirical relation between crack length and phase decay, `$a \ vs. \ \Delta\Phi$'. Finally, the results are discussed in Section \ref{Discussion}. The work in the paper shows the use of a very sensitive phase response as a practical and easy to implement new measurement technique to monitor very small crack length and growth.

\section{Experimental measurements}\label{EM}

This section explains the materials used for the composite laminates, the experimental setup designed to perform the High Frequency Fatigue Testing (HFFT) and their corresponding experimental procedures.

\subsection{Materials and experimental set up}

The composite material used in these experiments is T700/M21 Hexcel©, because of its widespread use in the industry and large amount of scientific data available. The laminates have the following lay up configuration: [90, 0-cut, 90-cut, 0, 90, 90, 0, 90, 0, 90]. The overall dimensions of the samples are 270x25x2.5mm (0.25mm of thickness per layer). The manufacturing of the laminates followed a standard vacuum bagging technique using autoclave. 

\begin{figure}[!ht]
	\centering
		\includegraphics[width=1\linewidth]{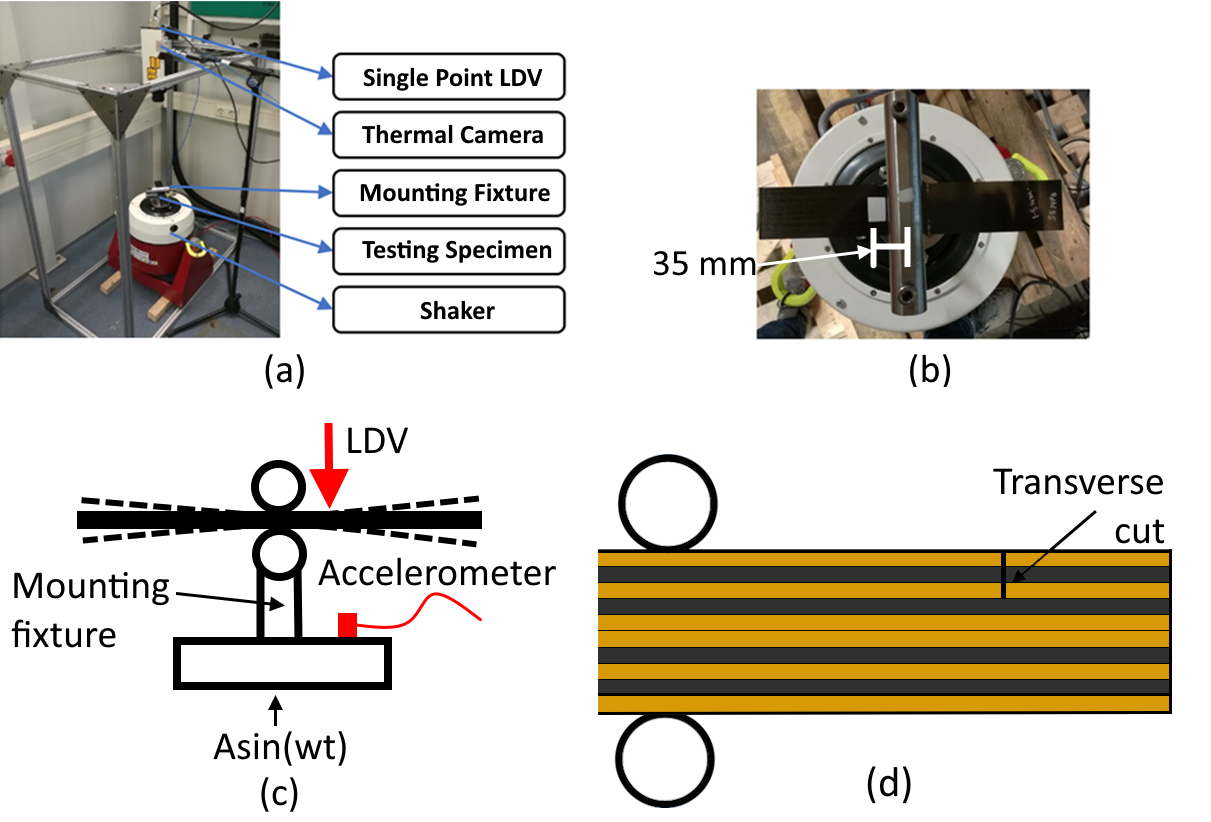}
	\caption{Experimental Set-up for HFFT. (a) Components. (b) Top View of sample and fixture. (c) Schematic of the measuring and excitation signals. (d) Schematic of the crack path.}
	\label{fig_1}
\end{figure}

The experimental set-up designed to perform the HFFT experiments is provided in \cref{fig_1}. In \cref{fig_1}(a), a shaker is exciting the testing specimen through a mounting fixture connecting the shaker to the middle width line of the specimen, with the objective of exciting the first bending vibration mode of the specimen. A thermal camera is used to check that self-heating behaviour does not occur, self-heating behaviour is to be avoided as its impact in the dynamics of composites is yet to be understood \cite{Voudouris2020}. To track the specimen displacement, an LDV is pointed to a location 30 mm from the mounting feature, as shown in \cref{fig_1}(b). \cref{fig_1}(c) show a schematic of the instrumentation which helps to track the shaker input, via an accelerometer held to the base of the mounting fixture and the LDV pointing to the top of the specimens to track its displacement.  It is important to emphasise that the vibration test creates a stress ratio R = -1, and therefore the crack formation will go through Mode-I and Mode-II, depending on the positive or negative oscillation part of the response. \cref{fig_1}(d) shows an illustration of the composite lay up, $90^o$-layers in orange and $0^o$-layers in blue and the defect used in the experiments. The initial defect is a transverse cut through 2 layers, which consequently spread in plane to the left and right of this initial cut.

\subsection{Experimental Procedures}

With common and economically accessible measurement equipment it is currently impossible to measure the evolution of embedded cracks in real time. Hence, two different procedures have been designed to perform fatigue testing on the samples. On one hand, a set of interrupted fatigue tests, aims to obtain crack lengths at the interruption moments of the fatigue testing, along with the phase decay and frequency of the fatigued component. The second set of tests, named long duration fatigue tests set, aims to give a notion of the uncertainty of the testing approach and the differences that raise with the interrupted tests. The interrupted and long duration test procedures are explains as follows.

The interrupted tests experimental procedure, as it names suggests, is and interrupted test on the same sample, approximately every 100.000 cycles. The overall procedure is as follows.

\begin{enumerate}
   \item FRF measurement of $(\phi(\omega),A(\omega))$,  to obtain initial phase, natural frequency of the first bending mode and damping. Here, a standard step sine sweep test is used.
   \item Fatigue testing at constant vibration amplitude and excitation frequency.
   \begin{enumerate}
     \item During the test a PID controller uses the measured displacement from the LDV and inputs a control voltage signal to the shaker targeting to the right strain severity level. The severity level used in the tests is 1.3 mm, which is good enough to produce a traceable fatigue response, out of the infinite life range and also not failing the material too fast.
     \item Once the strain level is found it is kept constant during the test, by tracking the displacement from the LDV continuous measurement. The strain level is found before the high cycle fatigue testing, by doing a simple calibration between the LDV measurement displacement and a strain gauge installed on the bottom side of the samples, below the laser point. To execute this calibration the samples are excited at the resonance frequency found in step 1, at progressively increased shaker voltage from 0.25 V input by four increments of $\Delta V$ = 0.25 V. This produces a straight line of displacement vs. strain.
     \item The phase between response (LDV) and reference (accelerometer) is recorded, see \cref{fig_1}(c). From this recordings a decaying phase can be observed.
     \item The frequency is periodically adjusted to keep the excitation at resonance as material strength decays from fatigue degradation \cite{Backe2015, Lorenzino2015}. 
    \end{enumerate}
   \item  The test is interrupted approximately every 100,000 cycles (hence the name interrupted tests).  
   \item A measurement is performed to obtain the crack length. The side of the samples is slightly polished and a microscopy observation is taken.
   \item The test is re-started, on the same sample, at step 2 of the procedure (from the last resonant frequency before interruption).
\end{enumerate}

The long duration tests follow the same steps 1 and 2 of the interrupted tests procedure, after this, without any interruption. Once the test is finished, the crack length is measured; to do this measurement the sample is cut and a standard CT-Scan is performed.

It is referred as phase `decay' because the input remains steady while the response slowly lags behind the excitation input more and more as the material strength decays. \cref{fig_1}(c) pictures this more clearly, it shows the sinusoidal excitation used in these tests from the shaker and into the fixture, which remains steady at constant amplitude and frequency, while the response from the LDV progressively lags behind. In other words, the phase lag: $\phi = \phi_{accel} - \phi_{LDV}$ increases in magnitude, because $\phi_{accel}$ is constant, but $\phi_{LDV}$ increases as the material strength is not enough to keep up with the speed of excitation, therefore lagging behind the input.

\section{Phase decay with number of cycles,  $\Delta\Phi(N)$.}\label{Res1}

This section presents the results of the experimental campaigns for the interrupted and long duration tests. 

\subsection{Interrupted Tests Results}

\begin{figure}[!ht]
	\centering
		\includegraphics[width=.7\linewidth]{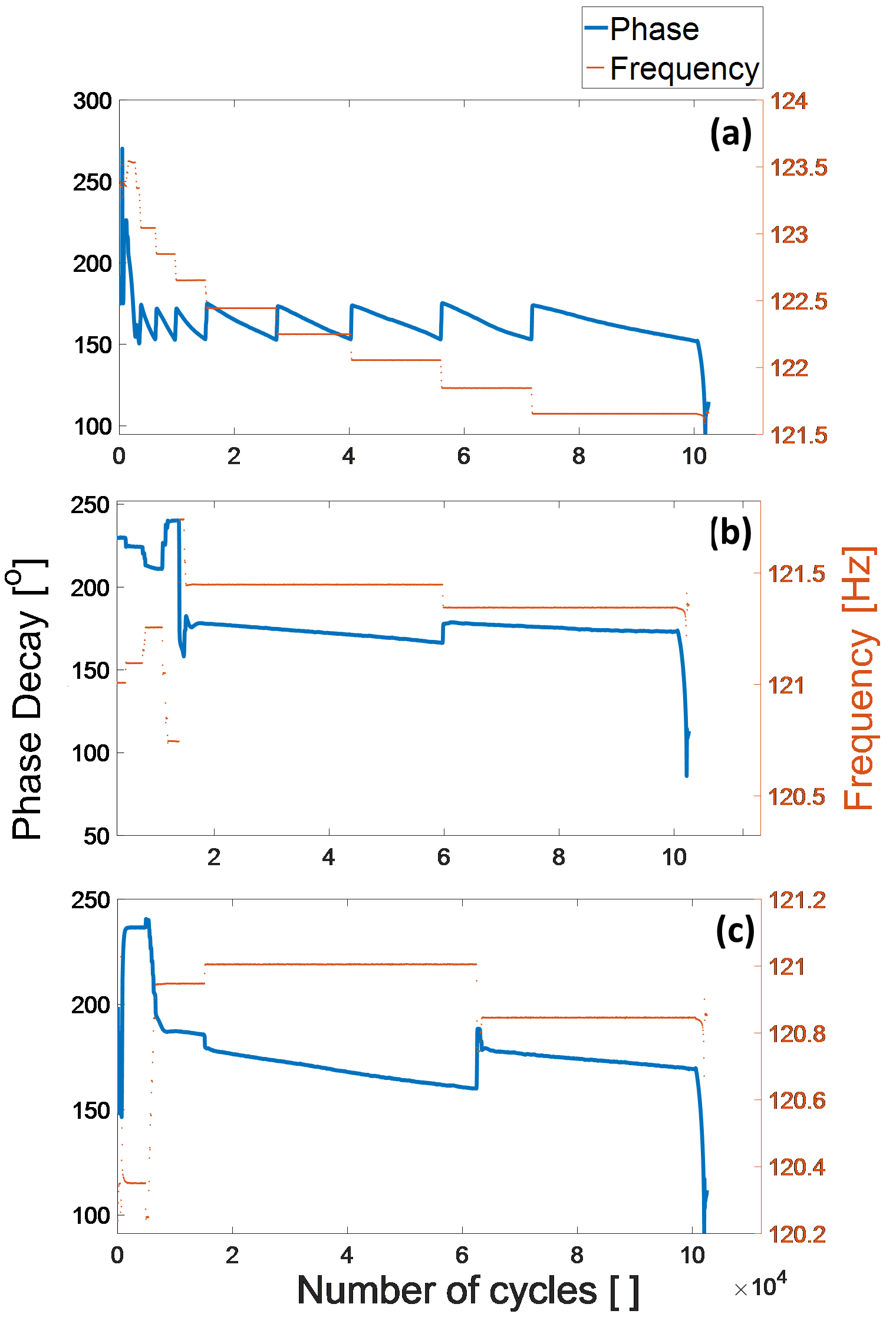}
	\caption{Phase response and frequency of excitation for the interrupted tests. (a) 0-100k cycles. (b) 100k-200k cycles. (c) 200k-300k cycles.}
	\label{fig_2}
\end{figure}

\cref{fig_2} shows the results of the three independent interrupted tests in terms of the excitation frequency and the phase decay. The frequency at which the system is excited, $\omega_e$, is also the first bending mode resonant frequency, exciting at this frequency demands less energy from the excitation system, in this case a shaker. The excitation frequency is in the neighbourhood of the natural frequency, because the natural frequency changes and the excitation frequency is fixed.

The response phase, $\phi$, starts from the phase of the undamaged case, found from the FRF in the step 2 described in the experimental procedures, and it progressively deteriorates with the damage evolution. 

At the beginning of the tests, there are a few cycles used for displacement amplitude adjustment, where the voltage applied into the shaker is changed to get to the desired strain level of the samples at the point measured by the LDV. Once the strain level severity is reached it is kept constant for the rest of the test. After this initial displacement adjustment, the phase decays continuously in segments, while the frequency remains constant. In consequence, each segment is kept at constant strain severity and frequency, allowing to attribute the phase decay to only a stiffness deterioration and a fatigue effect.

These segments are generated because the excitation frequency is adjusted in the control software, the resonant frequency of each sample changes due to fatigue stiffness degradation, and to keep the phase study within a reliable window, the excitation is not allowed to diverge to much. Keeping the excitation at resonance is of high importance, this assures remaining close to the most sensitive section of the phase response in the FRF, which corresponds to the highest magnitude of $d\phi \slash d\omega$ in an FRF. These segments in \cref{fig_2} are a consequence of adjusting down the excitation frequency $\omega_e$, to keep the excitation at resonant frequency and to take advantage of the sensitive part of the phase response. Hence, once an arbitrary phase threshold of 20° is reached, the excitation frequency is manually adjusted to match the resonance. From that point onwards, within each segment, the resonance changes from degradation, but the excitation frequency remains constant. 

It is worth noting that even though the natural and excitation frequencies progressively reduce, they remain considerably higher than the usual loading frequencies in fatigue standardised testing. High frequencies in fatigue tests may introduce self-heating, where viscous effects in the bulk material, or friction in delaminated areas, transform elastic energy into heat. However, the thermal camera revealed no observable changes in temperature at the top surface, which is 0.75 mm from the delamination plane between the 3rd and 4th layer, see \cref{fig_1}(d).

\begin{figure}[!ht]
	\centering
		\includegraphics[width=.7\linewidth]{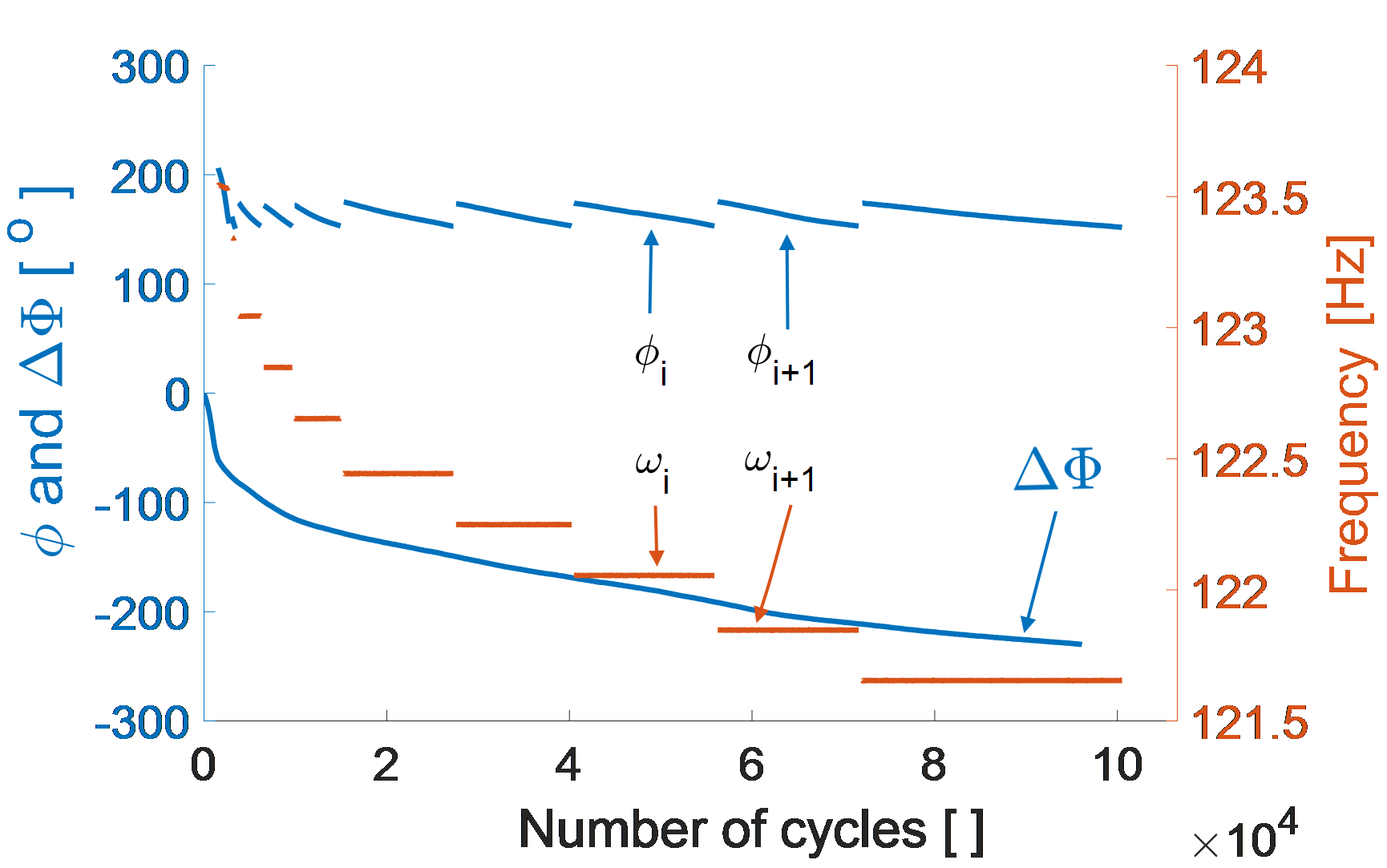}
	\caption{Phase decay segments and resonant frequency for the first 100k interrupted test and the concatenated phase decay curve.}
	\label{fig_3}
\end{figure}

\cref{fig_3} shows the data processing of the phase decay segments, using \cref{fig_2}(a) results as an example. These segments of phase decay, $\phi_i$, at constant frequency are concatenated to give a measurement of a total phase degradation,(say that this is not physical its an extra instrument) $\Delta\Phi = [\phi_1,\phi_2,\dots,\phi_n ]$. $\Delta\Phi$, is purposely started from 0 degrees, to indicate the phase difference with the initial undamaged case.  To concatenate the segments, the cycles corresponding to the initial displacement adjustment before reaching the natural frequency, and the frequency adjustment time laps, when jumping between $\omega_i$ and $\omega_{i+1}$, are ignored in the total number of cycles, as they do not contribute to fatigue at constant frequency. Then, the segments are put in sequential order, this is repeated for the individual segments in the three interrupted test every 100k cycles. The first segment of the first interrupted test, from 0-100k starts from a phase decay of 0 degrees, to indicate the pristine case starting point, and then the segments from the 100k-200k cycles and 200k-300k cycles are concatenated. \cref{fig_3} shows an example of this concatenation for the first interrupted test from 0-100k cycles.

\begin{figure}[!ht]
	\centering
		\includegraphics[width=.7\linewidth]{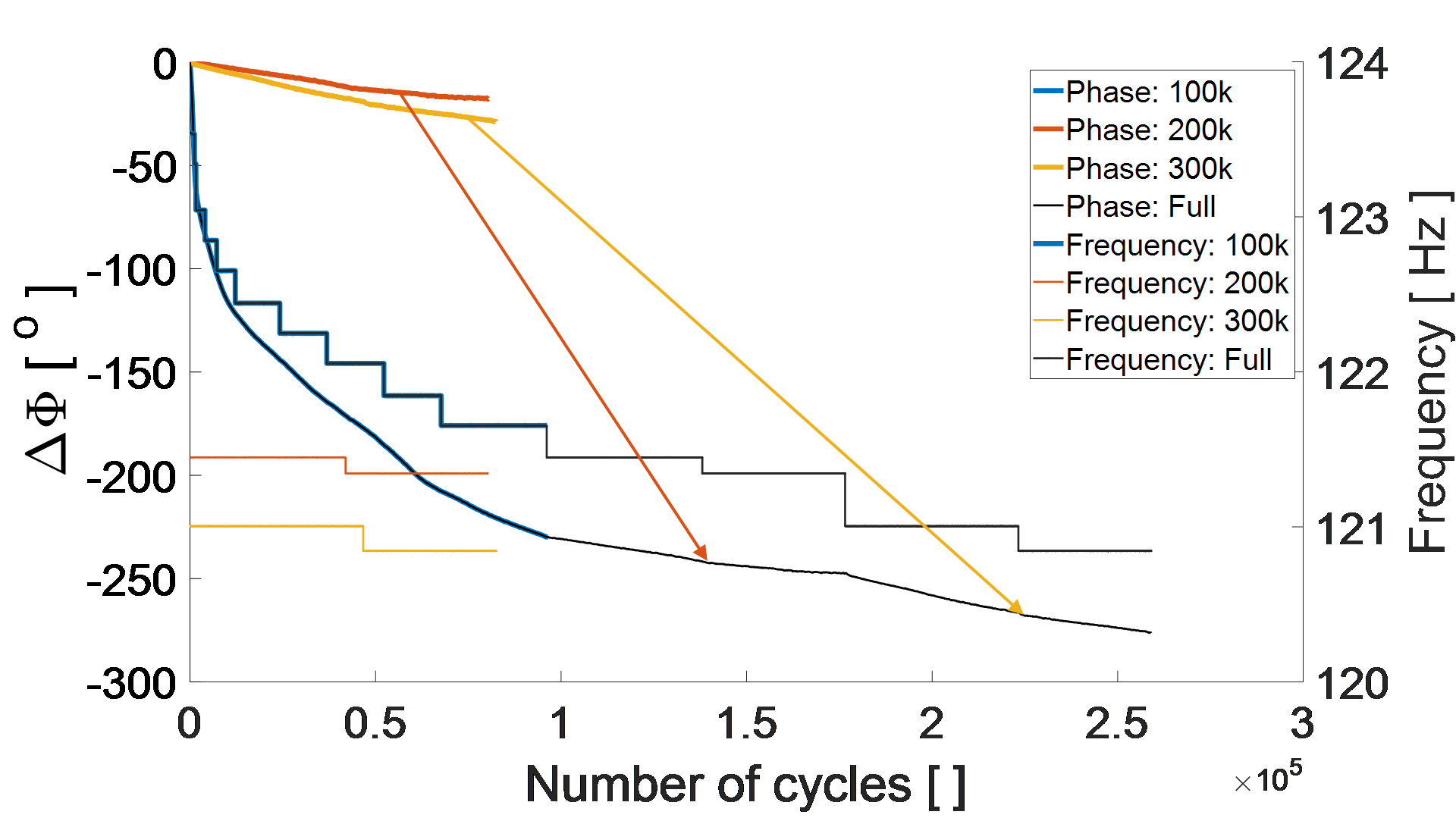}
	\caption{Total phase decay and resonant frequency change curves for the three interrupted tests and the full curve from 0 to 300k.}
	\label{fig_4}
\end{figure}

The full concatenation is merged in sequential order for the three interrupted tests in \cref{fig_4}. The method used for the final merge is similar to that of the individual segments,  integrating the total decay of the phase and frequencies from 0 to $\sim$300k cycles. 

\cref{fig_4} highlights that the phase decay is largely more sensitive to the degradation of the specimen than the resonant frequency. It can be seen, when looking at the vertical axes at the full phase decay and frequency curves, that the total phase decay is $\sim$ 280 degrees, whereas the resonant frequency change is only about 3 Hz. This shows the relevance of the phase as a signature parameter to appropriately track the fatigue of composite materials.

\begin{figure}[!ht]
	\centering
		\includegraphics[width=.7\linewidth]{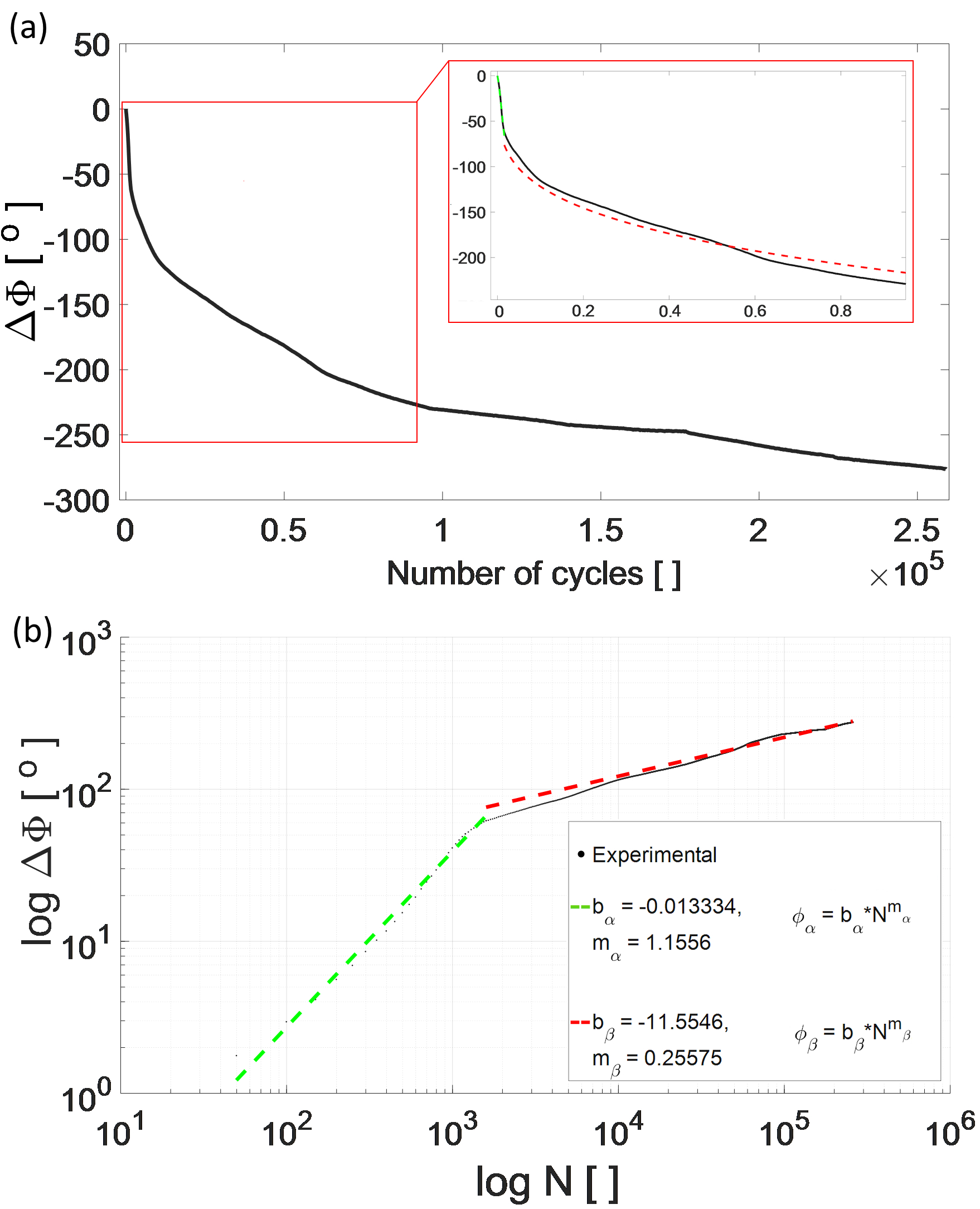}
	\caption{$\Delta\Phi(N)$ curve, from the concatenation of the phase decay results of interrupted tests. (a) Linear scale of the Phase decay, with a zoom into the inflexion point between zones $\alpha$ and $\beta$ of the power law fits. (b) Log-log scale with zones $\alpha$ and $\beta$ of the power law fits.}
	\label{fig_5}
\end{figure}

An experimental phase decay with number of cycles curve, `$\Delta\Phi(N)$' is established from this data. To obtain this empirical law, a logarithmic scale is used into the total phase decay to identify the inflexion points. Then, power law fits -in the linear scale- are imposed to the zones before and after the inflexion. \cref{fig_5} shows the full phase decay curve and its corresponding power law fit in log-log scale.

The $\Delta\Phi(N)$ curve shows that an empirical law, based on the phase response as a more sensitive signature of a component degradation, can be established for a composite. From this $\Delta\Phi(N)$ curve two fits are extracted, separated in Zones $\alpha$ and $\beta$. Where Zone-$\alpha$ shows a more rapid decay than Zone-$\beta$ with an inflexion point at $\sim$ 1600 cycles. The expression for zone-$\beta$, as it is the one of most interest, is written in \cref{eq1}, the zone sub-index has been dropped, here and onwards, for simplicity.

\begin{equation} \label{eq1}
\Delta\Phi = b \times N^m = -11.555 \times N^{0.256}
\end{equation} 

These zones have been named Zones-$\alpha\slash\beta$ to avoid the temptation to reference them to Regions I and II from a Paris-Erdogan Law. This is purposely avoided because the nature of the crack progression, transverse crack moving into two in-plane delamination directions, makes impossible to assure with certainty that the zones found here obey the characteristic micromechanic behaviour usually associated to the regions in a Paris-Erdogan law.

So far the approach has used a single test, in order to verify that this approach is valid more tests are required.

\subsection{Long Duration Tests Results}

To obtain a measure of the uncertainty, carried out by doing interrupted testing and the computation of phase decay as a degradation tool (as well as all the uncertainties associated by the samples themselves), long duration tests are executed and plotted along with the concatenated interrupted tests.

\begin{figure}[!ht]
	\centering
		\includegraphics[width=.7\linewidth]{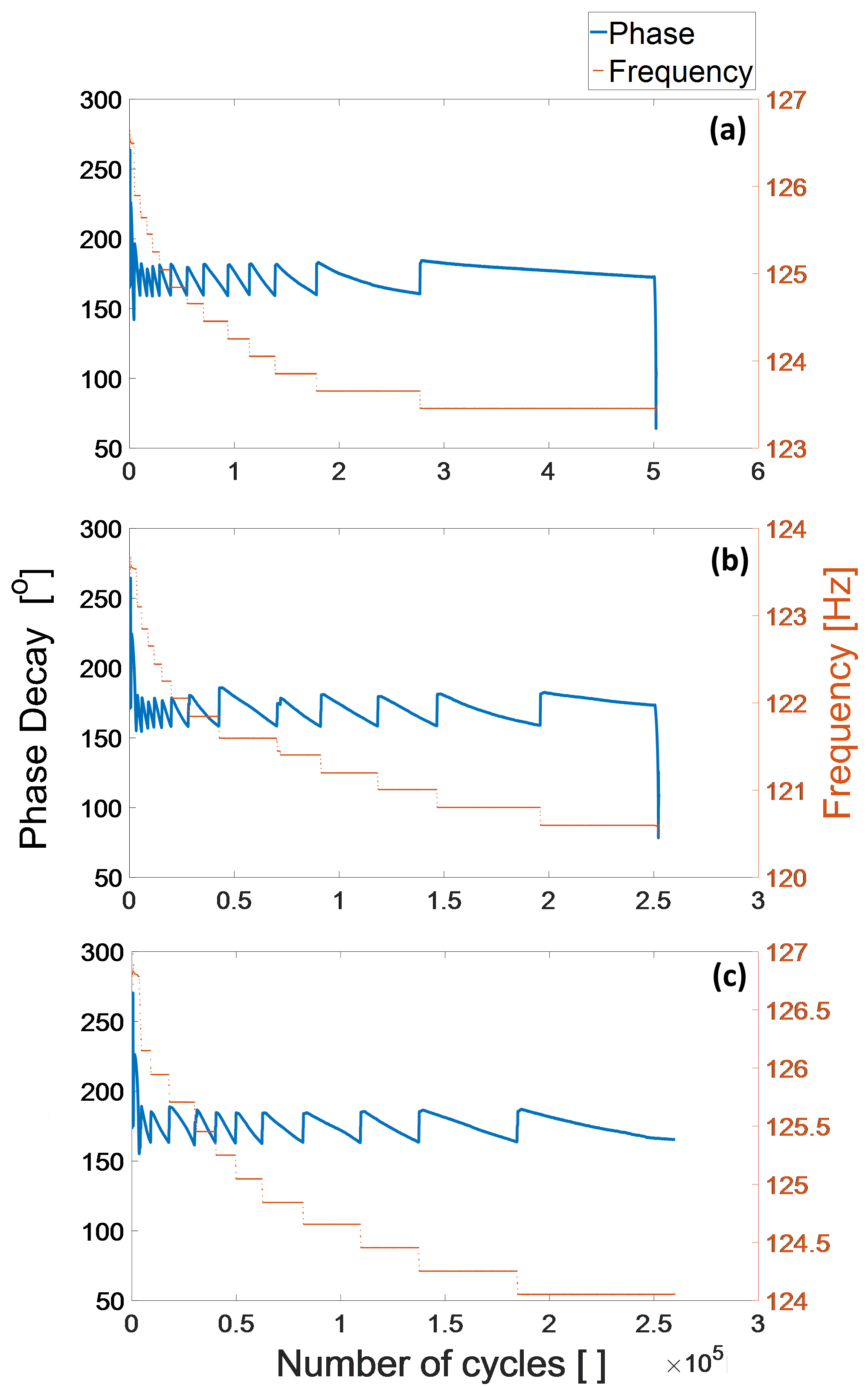}
	\caption{Phase response and frequency of excitation for the long duration tests. (a) Sample 1: 0-500k cycles. (b) Sample 2: 0-250k cycles. (c) Sample 3: 0-260k cycles.}
	\label{fig_6}
\end{figure}

\cref{fig_6} shows the raw results from the long duration tests in terms of its phase and frequency as fatigue evolves.

\begin{figure}[!ht]
	\centering
		\includegraphics[width=.7\linewidth]{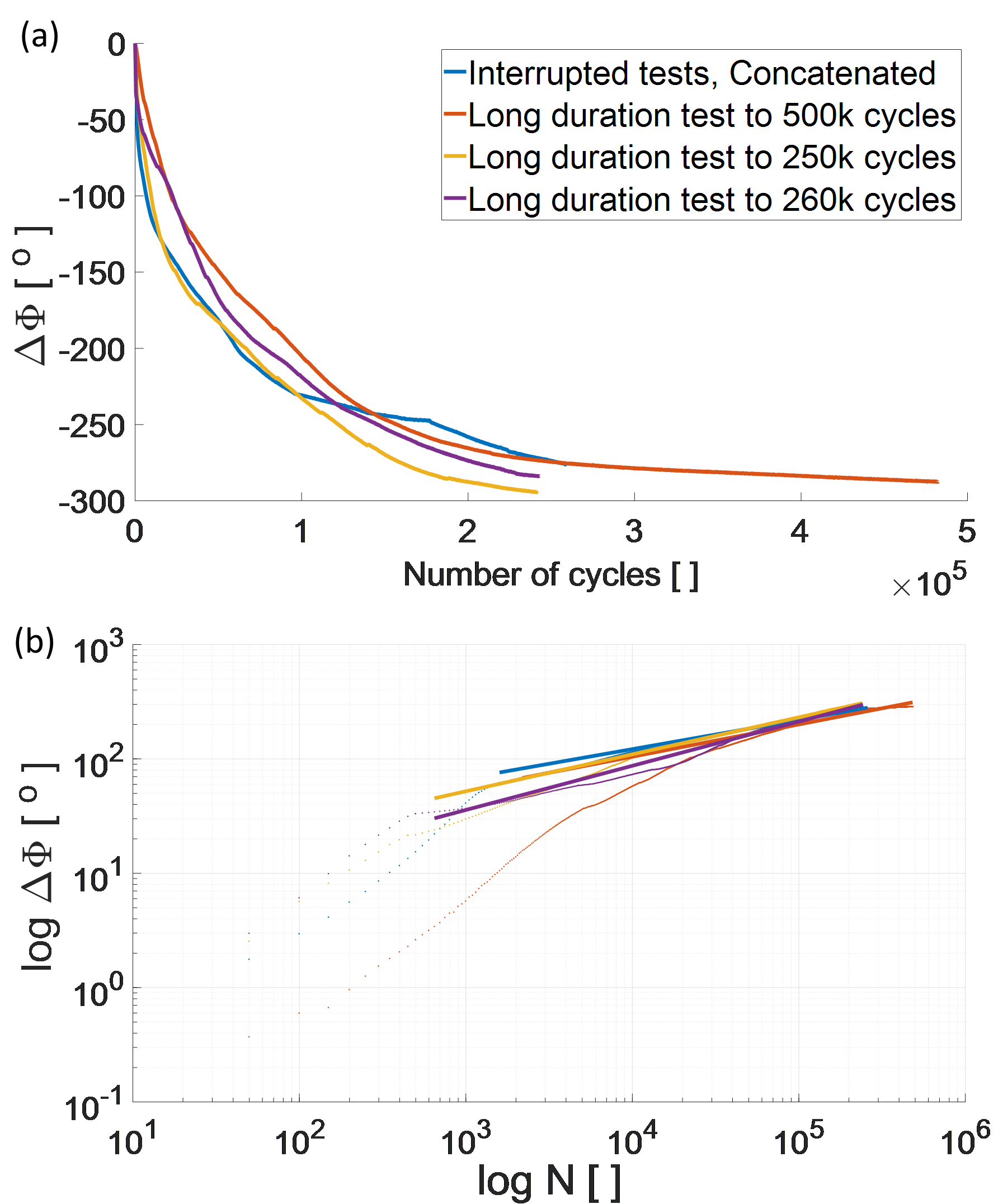}
	\caption{Phase response of the three long duration tests and concatenated interrupted tests. (a) Linear scale. (b) Log-log scale with Power law fits.}
	\label{fig_7}
\end{figure}

The processing of the results for these long duration tests follows the same procedure explained in \cref{fig_3} and \cref{fig_4}. The Long duration tests and the interrupted tests results are displayed in \cref{fig_7}.

It can be seen from \cref{fig_7} that the phase decay of the long duration tests shows a similar trend when compared with the interrupted tests. This increases the confidence in the proposed approach to obtain the $\Delta\Phi(N)$ curves. Table~\ref{table1} shows the coefficients of the power law fits for zone-$\beta$ for all the tests.

\begin{table}[]
    \caption{Power Law fit coefficients of the zone-$\beta$ of the long duration and concatenated interrupted tests.\label{table1}}
    \centering
    \begin{tabular}{lcc}
        		
         & $b_\beta$ & $m_\beta$ \\
        \midrule
        Interrupted tests (concatenated) & -11.555 & 0.256\\
        Long duration to 500k cycles & -7.799  & 0.281\\
        Long duration to 250k cycles & -5.587 & 0.323\\
        Long duration to 260k cycles & -2.484 & 0.386 \\ 
        \bottomrule
    \end{tabular}
\end{table}

However, even when following the same experimental procedure, they present some differences. These differences can be attributed to material and experimental uncertainties. Remarkably, the interrupted test follows quite well the tendency and the differences can be assigned to inconsistencies in the clamping of the specimens in the mounting fixture after taking them out in each interruption of the test for microscopy measurements.

\section{Crack progression and its relation to $\Delta\Phi$}\label{Res2}

This section explains how the interrupted tests are used to obtain an empirical relation between the phase decay and the crack length, such that the phase could be used as a reliable measure for crack progression.

Each of the three interrupted tests in the previous subsection finalised by measuring the crack length in the microscope. Therefore, each interrupted test has an associated outcome tuple of crack length, phase and number of cycles; ($a(N), \Delta\Phi(N)$). $\Delta\Phi$ is a function of the number of cycles, from the empirical law in \cref{eq1}, therefore the crack length can be linked to the phase decay. \cref{fig_7} shows that the interruption of a test does not modify considerably the slope of the Zone-$\beta$, therefore a relation between $\Delta\Phi$ and $a$ can be established relatively safely.

\begin{figure}[!ht]
	\centering
		\includegraphics[width=.7\linewidth]{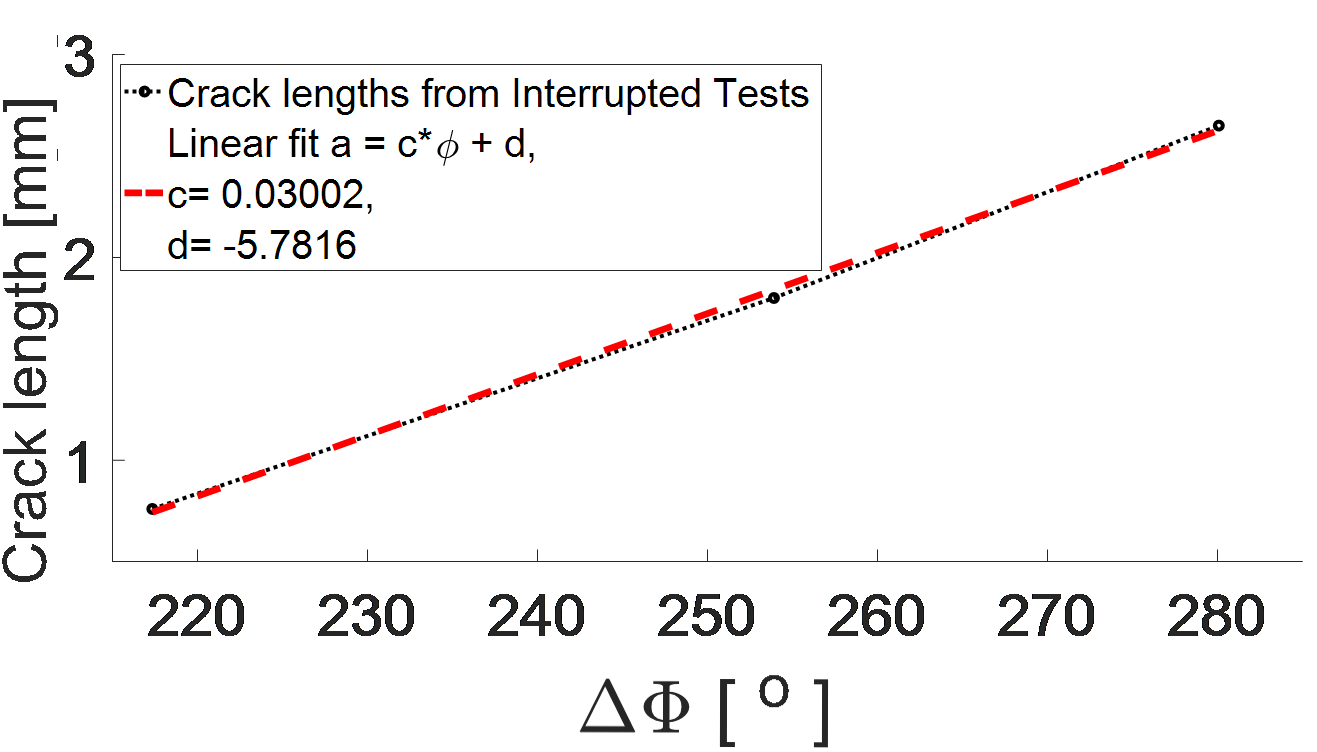}
	\caption{Crack length and phase decay empirical relation.}
	\label{fig_8}
\end{figure}

The experimental data of the measurement of the crack, the number of cycles at which the measurements were taken and the phase, obtained using \cref{eq1}, are displayed in Table~\ref{table2}. \cref{fig_8} displays these data with a linear fit, relating the crack length as a function of the phase decay.

\begin{table}[]
\centering
\caption{Crack length, number of cycles and phase at each interrupted test.\label{table2}}
\begin{tabular}{lccc}
Test &  N [ ] &  $a$ [mm]& $| \Delta\Phi [^o] |$ \\
\midrule
1st interruption & 96122 cycles  &  0.76 & 217.3\\
2nd interruption & 176549 cycles  & 1.8 & 253.9 \\
3rd interruption\\ (test end) & 259083 cycles & 2.65 & 280.1 \\
\bottomrule
\end{tabular}
\end{table}

The empirical relation, in \cref{eq2}, allows to write an expression relating the crack length as a direct function of the number of cycles, see \cref{eq3}.

\begin{equation} \label{eq2}
a = c \times \Delta\Phi +d = 0.003 \times \Delta\Phi -5.78
\end{equation} 

\begin{equation} \label{eq3}
a = c \times (b \times N ^ m) +d 
\end{equation} 

\begin{figure}[!ht]
	\centering
		\includegraphics[width=.7\linewidth]{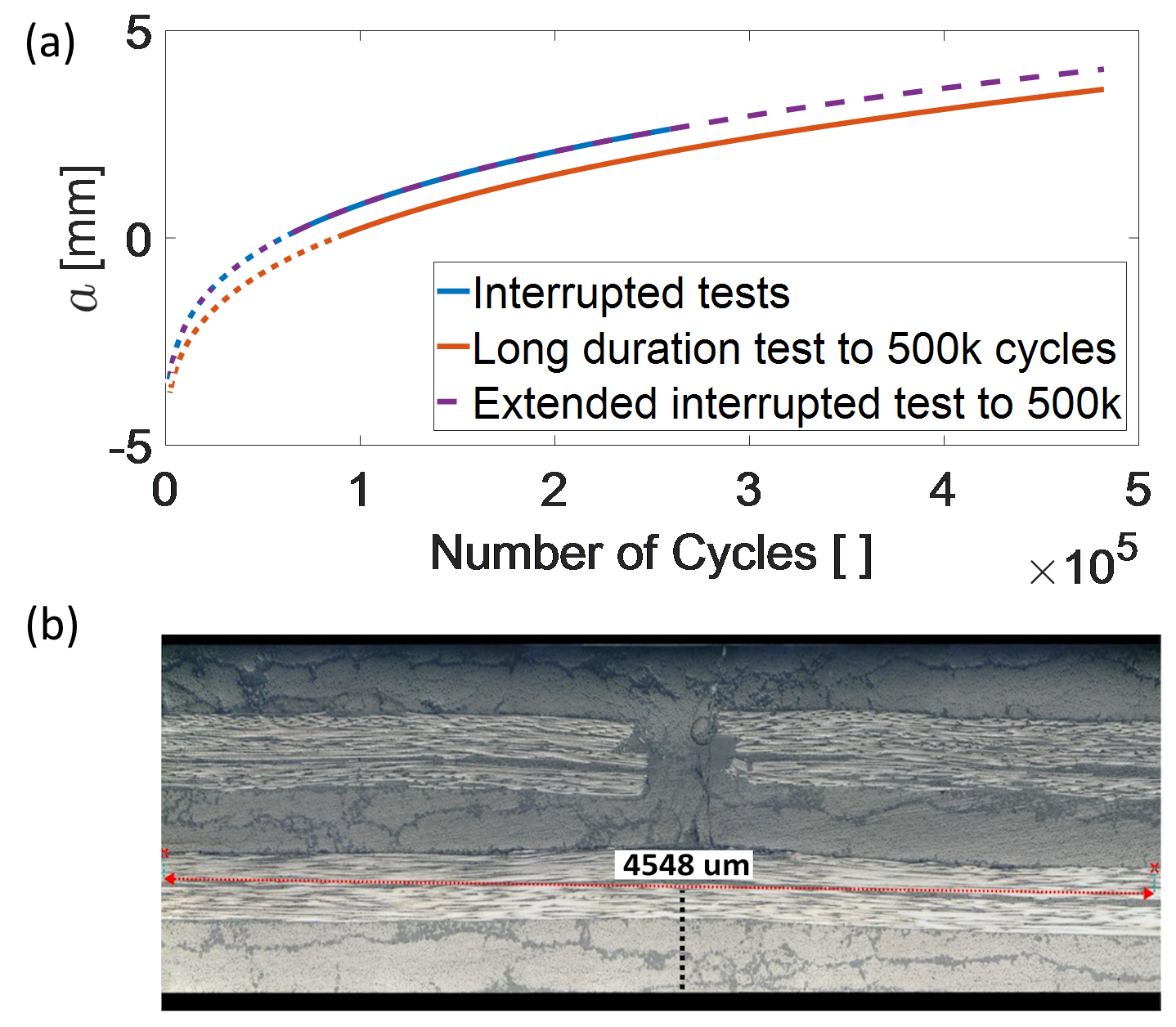}
	\caption{Empirical law for a(N) applied to the long duration tests results. (a) Empirical law using the coefficients extracted from the interrupted tests and extended to 500k, and using coefficients of both linear fit and power law of the long duration tests. (b) Measurement of the crack length of the long duration test.}
	\label{fig_10}
\end{figure}

To verify the applicability of these empirical laws, \cref{fig_10} shows the use of \cref{eq3} on the long duration experiment until 500k cycles, which allows the comparison between the empirical laws predictions against the experimentally measured crack length at the end of the test.

\cref{fig_10}(a) shows two curves corresponding to the interrupted tests (blue) and long duration experiments up to 500k cycles (red) (a third curved is added to project the prediction of the crack length from interrupted tests to 500k cycles for comparison against the long duration test, see purple curve). Interrupted and Long duration tests differ because, while both tests have characteristic phase decay, $\Delta\Phi(N)$, curves, only the interrupted testing approach allows to obtain the crack length measurements necessary to derive an empirical $a \ vs.\  \Delta\Phi$ curve, see the linear fit in \cref{fig_8}. Hence, the underlying assumption is that this linear fit, from the interrupted testing, applies to the long duration tests in similar specimens. To see if this assumption is reasonable, the coefficients from the linear fit on $a(\Delta\Phi)$ (the slope $c$ and intercept $d$) are used with the power law fits $\Delta\Phi(N)$ of both, the interrupted test coefficients ($b_\beta = -11.6$ and $m_\beta$ = 0.256) and 500k-long duration test coefficients ($b_\beta = -7.8$ and $m_\beta$ = 0.281), see Table~\ref{table1}.

The results show that the visible crack on the side of the 500k-long duration test extends 4.55 mm (\cref{fig_10}(b)), while the predictions using the approach described in the previous paragraph give 4.07 mm and 3.58 mm, respectively for the mixed interrupted/500k-long duration test coefficients and for only the interrupted test coefficients (\cref{fig_10}(a)). On the one hand the differences between predictions is attributed to material and implementations uncertainties inherent to manufacturing and also the interruption of the testing, on the other hand, the difference between the predictions and the measurements is believed to be due to the missing phase and crack-length threshold between Zone-$\alpha$ and Zone-$\beta$ giving noticeable negative values (see dotted lines for low number of cycles), which is discussed further in the next section.

With the analytical expression of $a(N)$, the crack propagation rate, $da \slash dN(N)$ is simply:

\begin{equation} \label{eq4}
\frac{da}{dN} = c \times | b | \times m \times N ^{m-1} = A \times N^ B
\end{equation} 

\begin{figure}[!ht]
	\centering
		\includegraphics[width=.7\linewidth]{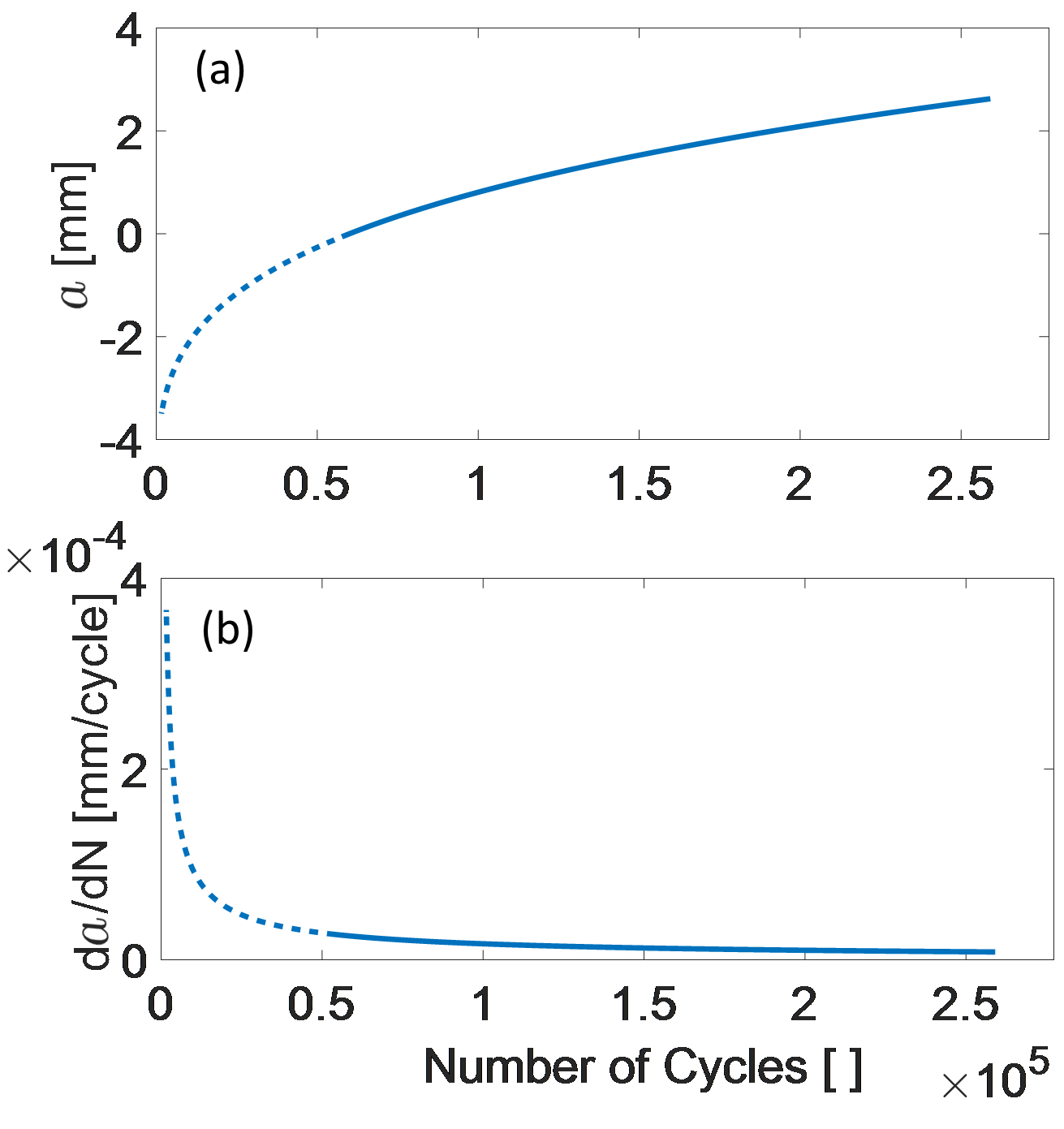}
	\caption{Empirical laws. (a) $a(N)$. (b) $da/dN(N)$}
	\label{fig_9}
\end{figure}

Where $ A = c \times | b | \times m = 0.0887\ [(mm/cycle)/(cycle^B)]  $ and $B = m-1 = -0.7442 $. \cref{fig_9} displays the crack length evolution $a(N)$ and $da \slash dN(N)$ from the interrupted test only, note that negative values of $a(N)$ have been presented with dotted lines.

\cref{fig_9}(b) demonstrates that the methodology described throughout this investigation can be used, provided the appropriate determination of Energy Release Rate or Stress intensity factors, to obtain Paris's Laws.

\section{Discussion}\label{Discussion}

\textbf{Test methodology and data processing}

The specimen is kept at a constant strain level, via controller on the derived displacement from the LDV signal. To avoid damage in excitation system, the excitation frequency is changed when an arbitrary threshold of phase decay is reached, which creates the phase decay segments in the raw data presented in \cref{fig_2} and \cref{fig_6}. Not doing so, requires large input voltages to keep the specimen vibrating at the desired strain level while not at resonance. The concatenation of these segments creates a representative total phase decay, which are physically meaningful as a strength decay signature, when they are at constant frequency and progressively decreasing in subsequent segments.

However, appropriately tracking the resonances presents a challenge in the control system when a test is interrupted. The removal and remounting of the specimens in the fixture introduces some uncertainties which complicates the measurement of the initial resonant frequency, ideally, the resonant frequency of the specimen before and after the interruption should be exactly the same. However, \cref{fig_2}(b-c) showed that this is not the case, from the somehow erratic behaviour in the phase in the initial cycles. In this publication these number of cycles of the initial frequency adjustment have been handled by simply ignoring them in the total curves, future versions of the control software improve the frequency tracking with a better controller.

Once the segments are concatenated and plotted in logarithmic scale, two characteristic zones are found Zone-$\alpha$ and Zone-$\beta$. Zone-$\alpha$ showing a more accentuated phase decay than Zone-$\beta$, more precisely, $m_{\alpha}\slash m_{\beta}=4.52$. The reasons behind this difference can be, for example, (i) Zones-$\alpha \slash \beta$ corresponds effectively to Regions I\slash II of a Paris-Erdogan Law, meaning that the micromechanics of the first $\sim$1600 cycles corresponds to the rapid progression of the cracks followed by a slower, and stable crack progression. (ii) Zones-$\alpha \slash \beta$ corresponds to either the left$\slash$right or right$\slash$left cracks growing independently from the transverse cut (see \cref{fig_1}(d)), where one direction stops growing before the other direction starts, or (iii) a combination of both (i) and (ii), it is plausible that both cracks grow simultaneously with each of them following a Paris-Erdogan behaviour. However, more experiments are required to identify these or new events, and to understand the physics of multi-directional cracks.

\textbf{Crack length$-\Delta\Phi$ relation}

A two-steps approach has been proposed to find an empirical relation of the progression of the crack with number of cycles. The first step builds the experimental phase decay, $\Delta\Phi$, from both interrupted and long duration tests. \cref{fig_7} increases the confidence in the results by giving phase decay results remarkably similar to the curve obtained from the interrupted tests. The second step, empirically relates the crack length from interrupted tests with the $\Delta\Phi$ found in the previous step, see \cref{fig_8}. Once the crack length is related to $\Delta\Phi(N)$, the crack length can be directly linked with the number of cycles for long duration tests, as shown in \cref{fig_9,,fig_10}. The predictions on the crack length propagation using this two-steps approach is always conservative, because it uses the empirical laws (log $\Delta\Phi$ vs. log N and $a$ vs. $\Delta\Phi$) corresponding to Zone-$\beta$, that is, it does not considers the total extension at the end of Zone-$\alpha$, that is partially the reason why the values of the crack length for a low number of cycles is negative. In other words, there is a a phase decay threshold, $\Delta\Phi_{th}$, that needs to be experimentally accounted for between Zone-$\alpha$ and Zone-$\beta$. The rapid speed of the crack progression in Zone-$\alpha$, $\sim$ 12.7 secs.(1600 cycles/125 Hz), along with the small dimensions of the studied cracks, are particularly inconvenient characteristics for interrupted testing, hence this work focuses on the characterisation of the crack progression of only the Zone-$\beta$, where a more stable progression and relatively larger crack dimensions facilitates the study and application of this methodology.

\section{Conclusions}\label{Conclusions}

This investigation has shown the experimental implementation and data processing of the HFFT technique using interrupted and long duration tests. The main results of this investigations are twofold. (i) The finding of a bi-linear law in the logarithmic scale of the phase decay of $\Delta\Phi(N)$, displaying a Zone-$\alpha$ of rapid more unstable crack growth followed by a slower and more stable crack progression zone, Zone-$\beta$. (ii) The description of an empirical crack progression law, $a(N)$, reached by a two-step process: (1) using the $\Delta\Phi(N)$ law mentioned in (i) and (2) a simple linear fit from the interrupted test crack length measurements relating $a \ vs. \ \Delta\Phi$. As a consequence, this empirical $a(N)$ law can be used to determine $da \slash dN(N)$ to find characteristic Paris-Erdogan Laws. 

The methodology of HFFT, applied together with interrupted and long duration tests, presents several advantages with respect to the standardised conventional methodologies of fatigue testing. (i) It is quicker to execute, because testing at resonance is normally faster than at low frequencies. It consequently has lower running costs because it takes less hrs. or day to run. (ii) Tracking the phase, or $\Delta\Phi$, to characterise material strength decay is two orders of magnitude more sensitive and therefore more useful than, f.ex., tracking the decay of the resonant frequency. (iii) The instrumentation required for this type of testing, LDVs and accelerometers, is accessible and standard within structural dynamics and fatigue laboratories which facilitates its implementation. Finally, (iv) the methodology in this investigation is not only applicable to determine crack progression, but it is also useful for the determination of characteristic Paris-Erdogan types of laws, albeit obtaining the corresponding Energy Release Ratios.

\section{Acknowledgements}\label{Acknowledgements}

The authors appreciate the funding from EU Horizon HEGEL Grant Agreement ID: 738130

\bibliographystyle{elsarticle-num}
\bibliography{My_bib}  

\begin{thebibliography}{10}
\expandafter\ifx\csname url\endcsname\relax
  \def\url#1{\texttt{#1}}\fi
\expandafter\ifx\csname urlprefix\endcsname\relax\def\urlprefix{URL }\fi
\expandafter\ifx\csname href\endcsname\relax
  \def\href#1#2{#2} \def\path#1{#1}\fi

\bibitem{Davila2020}
C.~G. Davila, C.~A. Rose, G.~B. Murri, W.~C. Jackson, W.~M. Johnston, \href{https://ntrs.nasa.gov/citations/20200003113}{Evaluation of fatigue damage accumulation functions for delamination initiation and propagation}, Tech. rep., NASA (2020).
\newline\urlprefix\url{https://ntrs.nasa.gov/citations/20200003113}

\bibitem{Allegri2013}
G.~Allegri, M.~R. Wisnom, S.~R. Hallett, A new semi-empirical law for variable stress-ratio and mixed-mode fatigue delamination growth, Composites Part A: Applied Science and Manufacturing 48 (2013) 192--200.
\newblock \href {http://dx.doi.org/10.1016/J.COMPOSITESA.2013.01.018} {\path{doi:10.1016/J.COMPOSITESA.2013.01.018}}.

\bibitem{Pascoe2013}
J.~A. Pascoe, R.~C. Alderliesten, R.~Benedictus, Methods for the prediction of fatigue delamination growth in composites and adhesive bonds – a critical review, Engineering Fracture Mechanics 112-113 (2013) 72--96.
\newblock \href {http://dx.doi.org/10.1016/J.ENGFRACMECH.2013.10.003} {\path{doi:10.1016/J.ENGFRACMECH.2013.10.003}}.

\bibitem{Murray2018}
B.~R. Murray, S.~Fonteyn, D.~Carrella-Payan, K.-A. Kalteremidou, A.~Cernescu, D.~V. Hemelrijck, L.~Pyl, Crack tip monitoring of mode i and mode ii delamination in cf/epoxies under static and dynamic loading conditions using digital image correlation, Proceedings 2018, Vol. 2, Page 389 2 (2018) 389.
\newblock \href {http://dx.doi.org/10.3390/ICEM18-05225} {\path{doi:10.3390/ICEM18-05225}}.

\bibitem{Obaid2006}
A.~A. Obaid, S.~Yarlagadda, M.~K. Yoon, N.~E. Hager, R.~C. Domszy, A time-domain reflectometry method for automated measurement of crack propagation in composites during mode i dcb testing, http://dx.doi.org/10.1177/0021998306061309 40 (2006) 2047--2066.
\newblock \href {http://dx.doi.org/10.1177/0021998306061309} {\path{doi:10.1177/0021998306061309}}.

\bibitem{Roh2019}
H.~D. Roh, S.~Y. Lee, E.~Jo, H.~Kim, W.~Ji, Y.~B. Park, Deformation and interlaminar crack propagation sensing in carbon fiber composites using electrical resistance measurement, Composite Structures 216 (2019) 142--150.
\newblock \href {http://dx.doi.org/10.1016/J.COMPSTRUCT.2019.02.100} {\path{doi:10.1016/J.COMPSTRUCT.2019.02.100}}.

\bibitem{Panin2017}
S.~V. Panin, V.~O. Chemezov, P.~S. Lyubutin, V.~V. Titkov, The algorithm of crack and crack tip coordinates detection in optical images during fatigue test, IOP Conference Series: Materials Science and Engineering 177 (2017) 012019.
\newblock \href {http://dx.doi.org/10.1088/1757-899X/177/1/012019} {\path{doi:10.1088/1757-899X/177/1/012019}}.

\bibitem{Maillet2013}
I.~Maillet, L.~Michel, G.~Rico, M.~Fressinet, Y.~Gourinat, A new test methodology based on structural resonance for mode i fatigue delamination growth in an unidirectional composite, Composite Structures 97 (2013) 353--362.
\newblock \href {http://dx.doi.org/10.1016/J.COMPSTRUCT.2012.10.024} {\path{doi:10.1016/J.COMPSTRUCT.2012.10.024}}.

\bibitem{Maillet2015}
I.~Maillet, L.~Michel, F.~Souric, Y.~Gourinat, Mode ii fatigue delamination growth characterization of a carbon/epoxy laminate at high frequency under vibration loading, Engineering Fracture Mechanics 149 (2015) 298--312.
\newblock \href {http://dx.doi.org/10.1016/J.ENGFRACMECH.2015.08.030} {\path{doi:10.1016/J.ENGFRACMECH.2015.08.030}}.

\bibitem{Magi2016}
F.~Magi, D.~D. Maio, I.~Sever, Damage initiation and structural degradation through resonance vibration: Application to composite laminates in fatigue, Composites Science and Technology 132 (2016) 47--56.
\newblock \href {http://dx.doi.org/10.1016/J.COMPSCITECH.2016.06.013} {\path{doi:10.1016/J.COMPSCITECH.2016.06.013}}.

\bibitem{Voudouris2020}
G.~Voudouris, D.~D. Maio, I.~A. Sever, Experimental fatigue behaviour of cfrp composites under vibration and thermal loading, International Journal of Fatigue 140 (2020) 105791.
\newblock \href {http://dx.doi.org/10.1016/J.IJFATIGUE.2020.105791} {\path{doi:10.1016/J.IJFATIGUE.2020.105791}}.

\bibitem{Backe2015}
D.~Backe, F.~Balle, D.~Eifler, Fatigue testing of cfrp in the very high cycle fatigue (vhcf) regime at ultrasonic frequencies, Composites Science and Technology 106 (2015) 93--99.
\newblock \href {http://dx.doi.org/10.1016/J.COMPSCITECH.2014.10.020} {\path{doi:10.1016/J.COMPSCITECH.2014.10.020}}.

\bibitem{Lorenzino2015}
P.~Lorenzino, A.~Navarro, The variation of resonance frequency in fatigue tests as a tool for in-situ identification of crack initiation and propagation, and for the determination of cracked areas, International Journal of Fatigue 70 (2015) 374--382.
\newblock \href {http://dx.doi.org/10.1016/J.IJFATIGUE.2014.08.002} {\path{doi:10.1016/J.IJFATIGUE.2014.08.002}}.

\end{thebibliography}






\end{document}